\documentclass[twocolumn,english,aps,pra,longbibliography,superscriptaddress, floatfix]{revtex4-1}
\pdfoutput=1
\usepackage[utf8]{inputenc}
\usepackage[english]{babel}
\usepackage[T1]{fontenc}
\usepackage{amsmath}
\usepackage{hyperref}
\usepackage{booktabs}
\usepackage{tikz}
\usepackage{lipsum}
\usepackage{bm}
\usepackage{algpseudocode}
\usepackage{algorithm}

\begin{document}
	
	\title{Adiabatic-Passage-Based Parameter Setting for Quantum Approximate Optimization Algorithm}
	
	\author{Mingyou Wu}
	\affiliation{School of Computer Science and Engineering, Southeast University, Nanjing {\rm 211189}, China}
	\author{Hanwu Chen}
	\email{hw\_chen@seu.edu.cn}
	\affiliation{School of Computer Science and Engineering, Southeast University, Nanjing {\rm 211189}, China}
	\affiliation{Key Laboratory of Computer Network and Information Integration (Southeast University), Ministry of Education,  Nanjing {\rm 211189}, China}
	
	

	\begin{abstract}
		The Quantum Approximate Optimization Algorithm (QAOA) exhibits significant potential for tackling combinatorial optimization problems. Despite its promise for near-term quantum devices, a major challenge in applying QAOA lies in the cost of circuit runs associated with parameter optimization. Existing methods for parameter setting generally incur at least a superlinear cost concerning the depth \textit{p} of QAOA. 
		In this study, we propose a novel adiabatic-passage-based parameter setting method that remarkably reduces the optimization cost, specifically when applied to the 3-SAT problem, to a sublinear level.  
		Beginning with an analysis of the random model of the specific problem, this method applies a problem-dependent preprocessing on the problem Hamiltonian analytically, effectively segregating the magnitude of parameters from the scale of the problem. Consequently, a problem-independent initialization is achieved without incurring any optimization cost or pre-computation. 
		Furthermore, the parameter space is adjusted based on the continuity of the optimal adiabatic passage, resulting in a reduction in the disparity of parameters between adjacent layers of QAOA. By leveraging this continuity, the cost to find quasi-optimal parameters is significantly reduced to a sublinear level.	
	\end{abstract}
	\maketitle
	
	\section{Introduction}\label{sec1}
	
	Quantum computation \cite{Nielsen2000} is an emerging computational model rooted in quantum mechanics that has demonstrated remarkable speedup in various computational problems. Algorithms such as the prime factorization algorithm \cite{Shor1994}, the Grover search \cite{Grover1997}, and the HHL algorithm \cite{Harrow2009} show provable exponential or polynomial accelerations compared to classical algorithms. Notably, despite the theoretical advantages inherent in quantum algorithms, the present computational landscape is constrained by the availability of only noisy intermediate-scale quantum (NISQ) devices \cite{Preskill2018}. The intrinsic noise in these devices limits their capacity to manifest a quantum advantage over classical computers when applying such algorithms. In the NISQ era, the imperative to develop noise-tolerant quantum algorithms becomes paramount. A focal point of concern revolves around the development of hybrid quantum-classical algorithms, which leverage the parameterized quantum circuits and adjust these parameters through classical optimization techniques  \cite{Cerezo2021VQA}. Prominent instances of such approaches include Quantum Neural Networks (QNN) \cite{Farhi2018QNN, Biamonte2017} tailored for machine learning tasks, the Variational Quantum Eigensolver (VQE)  \cite{Peruzzo2014VQE, Kandala2017VQE} designed for quantum chemistry calculations, and the Quantum Approximate Optimization Algorithm (QAOA) \cite{farhi2014QAOA}.
	
	QAOA is a promising approach for implementation on NISQ devices \cite{Moll2018,Harrigan2021}, showcasing significant computational potential in the domain of combinatorial optimization and receiving extensive attention in various aspects. In 2017, a near-optimal quantum circuit for the unstructured search was introduced by employing the Grover Hamiltonian as the problem Hamiltonian within the QAOA framework \cite{PRA2017}. This reveals that QAOA is, at a minimum, as efficient as the Grover search algorithm \cite{Grover1997}. Theoretical investigations into QAOA's computational capabilities have produced noteworthy results for specific instances \cite{PRA2018, Brandao2018, basso2021quantum, farhi2021QAOA}. Moreover, a spectrum of theoretical analyses have been conducted to elucidate the inherent computational power of QAOA \cite{farhi2019QAOA, barak2021, Streif2021}. The versatility of QAOA is exemplified by the proposal of several variants tailored to different problem classes \cite{Wang2020, bako2022, RUAN2023}. However, an intrinsic challenge arises in the form of the curse of dimensionality, particularly regarding the optimization cost associated with parameter setting. 
	
	The optimization cost of QAOA primarily arises from adjusting parameters for the $p$-layered parameterized quantum circuit. Analogous to quantum adiabatic algorithm (QAA) \cite{farhi2000QAA, farhi2001QAA} and VQE, QAOA defines the problem Hamiltonian $H_C$ and evolve the quantum state to the eigenstate with the maximal (or minimal) eigenvalue of $H_C$. Notably, QAOA diverges by alternately applying the problem Hamiltonian $H_C$ and mix Hamiltonian $H_B$ for $p$ iterations, with the evolution time parameterized by $ {\bm \gamma}=\left( \gamma_1, \gamma_2,\cdots, \gamma_p \right)$ and ${\bm \beta} = \left( \beta_1, \beta_2, \cdots, \beta_p \right)$. In the context of an NP-complete combinatorial optimization problem with $n$ variables, it is hard to maintain the depth of QAOA to be a constant or logarithm of $n$ \cite{farhi2020QAOA}. Specifically, when the depth $p= {\rm Poly}(n)$, the polynomial number of parameters leads to an exponential optimization cost for parameters setting. 
	Various techniques are employed to mitigate the optimization cost of parameter setting in QAOA, including classical optimization \cite{crooks2018performance}, machine learning algorithm \cite{verdon2019learning, Streif2020, Jain2022, Moussa2022} and other heuristic strategies \cite{PRX2020, boulebnane2021predicting, Wurtz2021Setting, Wurtz2021Warm, Sack2021, Tate2023}. Among these, the heuristic strategy in \cite{PRX2020} is notably representative, reducing the optimization cost to approximately polynomial on $p$, albeit remaining superlinear. This paper proposes an adiabatic-passage-based parameter setting method that further diminishes the cost, specifically when applied to 3-SAT problem, to sublinear concerning the depth $p$. Concretely, the cost exhibits logarithmic growth in limited simulation, showcasing the potential efficiency of the proposed method. Here, the optimization cost refers to the number of circuit runs required to find quasi-optimal parameters.
	
	This method commences with a rigorous analysis of the statistical property inherent in the given problem. Specifically, for the $k$-SAT decision problem, a reduction is applied to transform it into a max-$k$-SAT optimization concentrated exclusively on satisfiable instances. Subsequently, random $k$-SAT models on satisfiable instances are constructed, aligning with the random $k$-SAT model \cite{Achioptas2006}. The Hamiltonian for this random model is conceived as random variable, facilitating the acquisition of statistical properties pertaining to the problem Hamiltonian for random instances. Following this characterization, a problem-dependent preprocessing step is executed on the Hamiltonian, effectively decoupling the magnitude of parameters from the scale of the specific problem. 
	Consequently, a problem-independent initialization is formulated. Drawing inspiration from QAA, this initialization method provides judicious initial parameters for QAOA of full depth $p$ without incurring any optimization cost or pre-computation. This initialization method shares a similar form with the Trotterized Quantum Annealing (TQA) initialization \cite{Sack2021}, with the only difference lying in whether pre-computation is involved. Furthermore, this initialization can be adapted into a QAA-inspired parameter setting method with linear-varying parameters, akin to the INTERP heuristic strategy in \cite{PRX2020}, thus representing an intermediate iteration of the proposed method. 

	Furthermore, to better illustrate the landscape of the parameter space, the adiabatic passage (AP) is introduced \cite{SAP2016,STIRAP2017}. QAOA can be conceptualized as a parameterized adiabatic passage, wherein the optimization of the parameters represents the search for the optimal adiabatic passage. By leveraging the inherent continuity in the optimal adiabatic passage, an alternative parameter space is introduced to augment the smoothness of parameters, thereby mitigating the differences in parameters between adjacent layers. The adiabatic-passage-based (AP-based) parameter setting method effectively capitalizes on this parameter continuity, resulting in a substantial reduction in optimization costs. 
	To provide a comparative analysis, evaluations are conducted with the TQA initialization in \cite{Sack2021} and the heuristic strategies in \cite{PRX2020}. In simulations of the 3-SAT problem, the optimization costs required for obtaining optimally linear-varying parameters for the INTERP is on the order of ${\mathcal O} (p)$, while the QAA-inspired and TQA method maintains constant costs despite the increasing in depth $p$. RRegarding the quasi-optimal parameters, the optimization cost of FOURIER exhibits a growth rate slightly exceeding ${\mathcal O} (p^2)$, while the TQA method experiences an increase slightly faster than linear. Notably, the AP-based method further diminishes this cost, manifesting a sublinear level. Specifically, it exhibits an approximately logarithmic relationship with depth $p$ in limited simulations. This reduction in optimization costs is attributed to the inherent continuity of the optimal adiabatic passage, resulting in only a logarithmic outer loop in the AP-based method which provides the potential for exponential acceleration.
	
	The subsequent sections of this paper are organized as follows. Section \ref{sec_preli} introduces some preliminaries to establish foundational concepts. In Section \ref{sec_preproc}, we delve into the statistical properties of the random $k$-SAT model. Based on this analysis, the Hamiltonian of QAOA undergoes preprocessing. Within Section \ref{sec_preproc}, according to the preprocessing, we initially introduce a QAA-inspired parameter initialization, which can be further adapted into a QAA-inspired parameter setting method with linear parameters. Additionally, we explore parameter continuity through an analysis of the optimal adiabatic passage, introducing an alternative parameter space with improved continuity. Building upon this analysis, we propose the AP-based parameter setting method. Section \ref{sec_comp_pref} compares the methods with the TQA initialization \cite{Sack2021} and heuristic strategies proposed in \cite{PRX2020} by simulations on 3-SAT, providing an experimental result about the potential exponential speed-up in parameter adjusting. In Section \ref{sec_discussion}, we conduct a discussion on the universal applicability of this method to general combinatorial optimization problems. Finally, Section \ref{sec_conclu} concludes this paper and suggests avenues for further research.

	\section{Preliminaries}\label{sec_preli}
	
	\subsection{Random $k$-SAT problem} \label{subsec_k_SAT}
	
	The Boolean satisfiability (SAT) is a well-known NP-complete problem of determining whether there exists an interpretation that satisfies a given Boolean formula. In the context of the $k$-SAT problem, the Boolean formula is confined to conjunctive normal form, where each clause is constrained to at most $k$ literals. In this paper, we further impose a restriction to exactly $k$ literals for each clause. The NP-completeness would be maintained as long as $k \ge 3$. This $k$-SAT decision problem can be reformulated to an optimization version, max-$k$-SAT, by defining the goal function as $C(x)=\sum_\alpha C_\alpha(x)$, where $\alpha$ denotes a clause, and $C_\alpha(x)$ represents the characteristic function of $\alpha$. Specifically, $C_\alpha(x)=1$ if $\alpha$ is satisfied by assignment $x$, and $C_\alpha(x)=0$ otherwise. max-$k$-SAT seeks to identify an assignment $t$ that maximizes $C(x)$. It is noteworthy that for max-$k$-SAT, the NP-completeness persists for $k\ge2$.
	
	\begin{figure}[!t]
		\centering
		\begin{minipage}[c]{0.45\textwidth}
			{\includegraphics[width=1\linewidth]{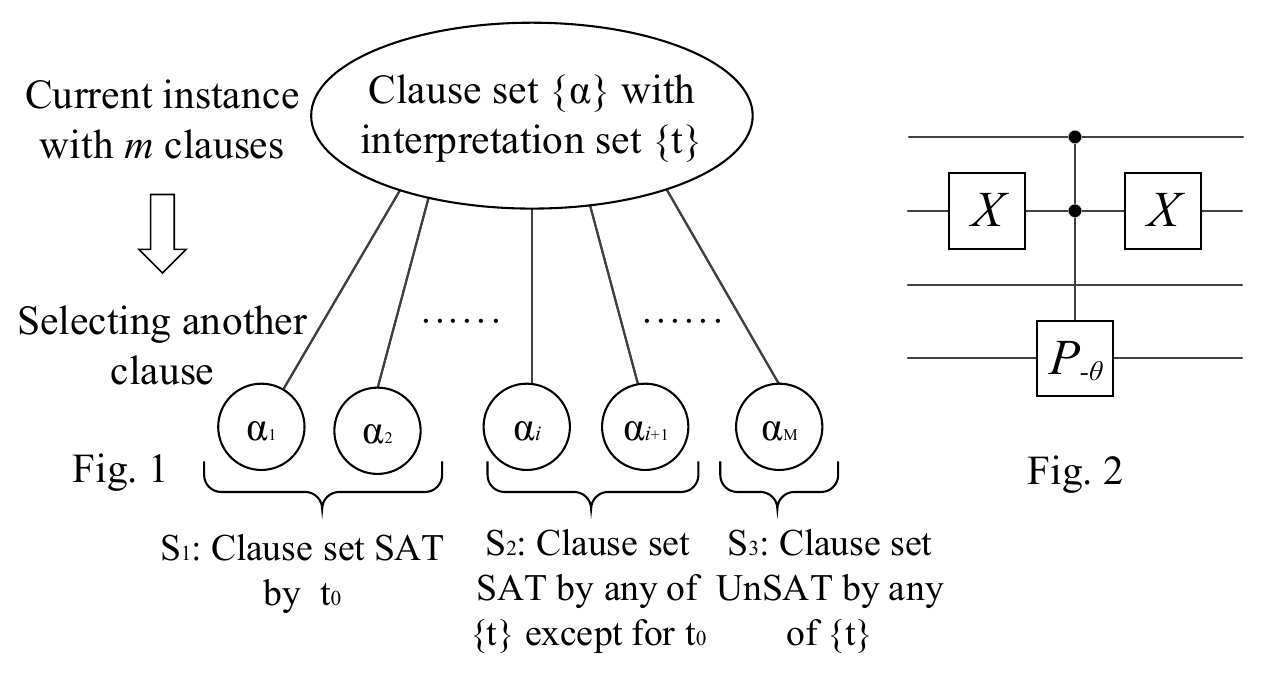}}
		\end{minipage}
		\footnotesize
		\caption{The process of selecting a clause when generating a random $k$-SAT instance. Let $I\in U_s$ and has interpretation set $\{t\}$. In the case of $F(n,m,k)$, when selecting the next clause, it randomly chooses any clause from $S_1+S_2+S_3$. In contrast, $F_s(n,m,k)$ selectively picks clauses that are satisfiable for any $t\in \{t\}$, specifically from $S_1+S_2$. Meanwhile, $F_f(n,m,k)$ exclusively selects clauses from $S_1$. When $\{t\}=\{t_0\}$, $F_s(n,m,k)$ coincides with $F_f(n,m,k)$, and with an increase in $m$, it converges to $F_f(n,m,k)$.}
		\label{fig_clause}
		\footnotesize
		\caption{An illustrative example of quantum circuit for $e^{i\theta H_\alpha}$, where $\alpha = \lnot x_1 \lor x_2 \lor \lnot x_4$. This clause is represented as $\alpha = (-1, 2, -4)$, corresponding to the Hamiltonian term $H_\alpha = -p_{\bar\alpha}$, where $\bar\alpha=(1, -2, 4)$. This implies that the 1st, 2nd, and 4th qubits (from top to bottom in the figure) are occupied, and an additional pair of $X$ gate is applied to the 2nd qubit. In this circuit, the 4th qubit serves as the controlled qubit; however, it is notable that the controlled qubit can be any of the involved qubits, with the rest qubits acting as the control qubits.}
		\label{fig_U1gate}
	\end{figure}
	
	The $k$-SAT decision problem can be reduced to max-$k$-SAT focusing exclusively on Satisfiable instances, denoted as \textbf{max-$k$-SSAT} in this paper. The k-SAT problem is polynomially solvable if max-$k$-SSAT exhibits polynomial solvability. A concise proof is presented as follows: The complete set $U$ of $k$-SAT is divided into two subset, $U_s$ comprising all satisfiable instances and $U_u$ containing the unsatisfiable ones. Assuming an algorithm $A$ can solve max-$k$-SSAT in ${\mathcal O}(f(n))$ time, another algorithm $A'$ tailored for $k$-SAT can be devised. Algorithm $A'$ accepts all instance $I\in U$ and processes $I$ by $A$. If running time exceeds ${\mathcal O}(f(n))$, the algorithm terminates and outputs a random result. For any $I$, $A'$ halts within ${\mathcal O}(f(n))$ time, yielding result $x$. In case when $I\in U_s$, $x$ serves as the interpretation $t$ and satisfies the Boolean formula; otherwise, $x$ is a random value and necessarily unsatisfiable. Through verification, the satisfiability can be determined. The verification cost dose not surpass ${\mathcal O}(f(n))$, maintaining an overall complexity of ${\mathcal O}(f(n))$.
	
	The random $k$-SAT model, denoted as $F(n,m,k)$, is widely employed in SAT problem research \cite{Achioptas2006}. Given the reduction of $k$-SAT to max-$k$-SSAT, random models exclusively on $U_s$ can be designed. $F(n,m,k)$ generates a $k$-SAT instance on $n$ variables by uniformly, independently, and with replacement selecting $m$ clauses from the entire set of $2^kC_n^k$ possible clauses. Building upon this foundation, two types of random models on $U_s$ are introduced. $F_s(n,m,k)$ serves as a direct extension from $F(n,m,k)$ by selectively choosing clauses while maintaining satisfiability. However, $F_s(n,m,k)$ presents theoretical challenges for analysis due to the varying optional clause set with $m$. An alternative random model, $F_f(n,m,k)$, is proposed as an approximation of $F_s(n,m,k)$. In this model, a pre-fix interpretation $t_0$ is randomly provided, and only clauses satisfied by $t_0$ are selected. If the generating process is conceptualized as a tree, the distinctions among these three random models are visually presented in Figure \ref{fig_clause}. In this paper, the focus lies primarily on $F_s(n,m,k)$ and $F_f(n,m,k)$.
	
	\subsection{Quantum adiabatic computation}\label{subsec_QAC}
	
	In the quantum system, the evolution of the state $\left| \psi(t) \right>$ is dictated by the Schr{\"o}dinger equation
	\begin{equation}
		i \frac{{\rm d}}{{\rm d} t} \left| \psi(t) \right> = H(t) \left| \psi(t) \right>,
	\end{equation}
	where $H(t)$ represents the system Hamiltonian. If $H(t)$ is time-dependent and $\left| \varphi(t) \right>$ is initially in the ground state of $H(0)$, the adiabatic theorem asserts that $\left| \psi(t) \right>$ will persist in the ground state of $H(t)$ as long as $H(t)$ varies sufficiently slowly.  
	
	Quantum adiabatic computation \cite{RMP2018AQC}  leverages the adiabatic theorem to solve for the ground state of a given problem Hamiltonian $H_C$. By designing the system Hamiltonian as 
	\begin{equation}
		H(s) = sH_C +(1-s)H_B, 
	\end{equation}
	and preparing the initial state $\left| \psi(0) \right>$ in the ground state of $H_B$, the state evolves to the ground state of $H_C$ as $s$ slowly varies $s$ from 0 to 1. The evolution time $T$ for $H(s)$ should satisfies
	\begin{equation}
		T \gg \frac{\varepsilon_0}{g^{2}_{0}},
	\end{equation} 
	where $g_{0}$ is the minimum of energy gap between the ground state $\psi_1(s)$ and the first excited state $\psi_2(s)$ of $H(s)$. Denoting $g(s)$ as the gap between $\psi_1(s)$ and $\psi_2(s)$, $g_0=\min_s{g(s)}$. Additionally, $\varepsilon_0$ is determined by the maximum of the derivative of $H(s)$, given by 
	\begin{equation}\label{eq_T_varepsilon}
		\varepsilon_0 = \max_s{\left < \psi_1(s) \left| \frac{\rm d}{{\rm d}s} H(s) \right| \psi_2(s) \right >}. 
	\end{equation}
	
	In simulation, the mapping from $s\in \left[0, 1\right]$ to $t\in \left[0,T\right]$ must be determined. The Quantum Adiabatic Algorithm (QAA) \cite{farhi2000QAA, farhi2001QAA} employs a straightforward approach by utilizing a linearly varying system Hamiltonian
	\begin{equation}\label{eq_QAA}
		H(t)=(1-\frac{t}{T})H_B + \frac{t}{T}H_C,
	\end{equation}
	where $0\le t \le T$. Here $H_B$ represents the transverse field $\sum_j \sigma_x^{(j)}$ and the initial state is the superposition state $\left| + \right>^{\otimes n}$, where $ \sigma_x^{(j)}$ denotes $ \sigma_x$ on the $j$-th qubit. Consequently, $\frac{\rm d}{{\rm d}t} H(t)$ is invariant and the required evolution time of QAA is $\mathcal{O}(g_0^{-2})$.

	\subsection{Quantum approximate optimization algorithm}\label{subsec_QAOA}
	
	Combinatorial optimization is a pervasive problem with applications across various domains \cite{COP2021}. Such problems typically involve a specific goal function $C(x)$, where the aim is to find an optimal target $t$ from a finite set $\{x\}$ that maximizes $C(x)$. The quantum approximate optimization algorithm is designed to address these combinatorial optimization problems. It formulates the problem Hamiltonian as ${{H}_{C}}\left| x \right> =C(x)\left| x \right>$, and seeks the target state $\left| t \right>$ with maximal energy using a layered variational quantum circuit expressed as
	\begin{equation}\label{eq_QAOA}
		\left| \bm{\gamma}, \bm{\beta}  \right> =\prod_{d=1}^{p} \left(e^{-i\beta_d H_B}e^{-i\gamma_d H_C}\right){{\left| + \right>}^{\otimes n}}.
	\end{equation}
	Here mix Hamiltonian $H_B$ is transverse field $\sum_j \sigma_x^{(j)}$, and $p$ represents the layer depth. The variational parameter $(\bm{\gamma}, \bm{\beta})$ are optimized to maximize the expectation $\left\langle {{H}_{C}} \right> =\left\langle  \bm{\gamma}, \bm{\beta}  \right|{{H}_{C}}\left|  \bm{\gamma}, \bm{\beta} \right\rangle$.
	
	QAOA is inspired by the Trotterization of adiabatic quantum computation, thereby sharing a similar framework with QAA. In simulation, the time-dependent Hamiltonians $H(t)$ can be Trotterized as a sequence of time-independent Hamiltonians using a sufficiently small $\Delta t$. Denoting $p=T/\Delta t$, the Trotter formula enables the approximate simulation of $H(t)$ for time $T$ through iterations of evolutions 
	\begin{equation}
		e^{iH(d\Delta t) }\approx e^{-i(1-\frac{d\Delta t}{T})H_B\Delta t}e^{-i\frac{d\Delta t}{T}H_C\Delta t},
	\end{equation}
	where $1\le d \le p$. This approximate simulation mirrors the framework of QAOA, differing only in the evolution time of Hamiltonians that are adopted as parameters in QAOA. 
	
	Generally, the diagonal $H_C$ can always be decomposed by the Walsh operator~\cite{Welch2014}, and for $k$-SAT, it can be further decomposed into the form of the Ising model \cite{Lucas2014}. To analyze the statistical properties of $H_C$, another problem-oriented operator for $H_C$ of $k$-SAT is provided. A Boolean conjunctive $c = (\lnot) x_{c_1} \land (\lnot)x_{c_2} \land \cdots (\lnot)x_{c_k} $ can be denoted as $(\pm c_1,\pm c_2,\cdots,\pm c_k)$, where $+c_t$ stands for $x_{c_t}$ and $-c_t$ for $\lnot x_{c_t}$. The corresponding problem Hamiltonian of $c$ is written as 
	\begin{equation}
		{{p}_{c}}=\sigma_{z^\mp}^{(c_1)} \otimes \sigma_{z^\mp}^{(c_2)} \otimes \cdots \otimes \sigma_{z^\mp}^{(c_k)},
	\end{equation}
	where $\pm c_j$ corresponds to $\sigma_{z^\mp}^{(c_j)}$. $\sigma_{z^\pm}^{(c_j)}$ is $\sigma_{z^\pm}$ on the $c_j$-th bit, where $\sigma_{z^\pm} = \frac{1}{2}(I\pm\sigma_z)$ are the corresponding components of $\sigma_z$ on $\left| 0 \right>$ and $\left| 1 \right>$, in matrix form as 
	\begin{equation}
		\sigma_{z^+} = \begin{bmatrix} 1 & 0 \\ 0 & 0 \end{bmatrix}, \quad
		\sigma_{z^-} = \begin{bmatrix} 0 & 0 \\ 0 & 1 \end{bmatrix}.
	\end{equation}

	For $k$-SAT, every clause $\alpha$ is a Boolean disjunctive, presented as $\alpha = (\lnot) a_{c_1} \lor (\lnot)a_{c_2} \lor \cdots (\lnot)a_{c_k}$. Denoting $\bar{\alpha}=(\mp a_1,\mp a_2,\ldots, \mp a_k)$, $H_\alpha = -p_{\bar{\alpha}}$. Consequently, the problem Hamiltonian of $k$-SAT is written as
	\begin{equation}\label{eq_HC}
		{{H}_{C}}=-\sum\limits_{\alpha } {p}_{\bar{\alpha }}.
	\end{equation}
	This Hamiltonian reduces to the Ising model by bringing in $\sigma_{z^\pm} = \frac{1}{2}(I\pm\sigma_z)$ \cite{Lucas2014}. The evolution $e^{i\theta H_\alpha}$ is a $(k-1)$-control phase gate, with a circuit complexity of ${\mathcal O}(k)$~\cite{Barenco1995, Saeedi2013}. Here the phase gate refers to the gate $e^{i\theta\sigma_{z^-}}$, denoted as $P_\theta$ in this paper, in matrix form as
	\begin{equation}
		P_\theta = \begin{bmatrix} 1 & 0 \\ 0 & e^{i\theta} \end{bmatrix}.
	\end{equation}
	In Figure \ref{fig_U1gate}, an example is presented.

	\section{Hamiltonian preprocessing of QAOA based on random $k$-SAT model}\label{sec_preproc}
	
	Based on the random $k$-SAT model $F_f(n,m,k)$ and the Hamiltonian decomposition with $P$ operator, as presented in Section \ref{sec_preli}, an analysis of the statistical properties of problem Hamiltonian $H_C$ for $k$-SAT can be conducted. Examining the process of $F_f(n,m,k)$ in clauses selection, since every $\alpha$ is randomly chosen, the diagonal Hamiltonian $H_\alpha$ can be viewed as a random vector, and its eigenvalues $E_{\alpha, x}$ at $\left| x \right>$ also become random variables. Due to the property that $F_f(n,m,k)$ selectively picks clauses satisfiable by $t$, the mean of ${{E}_{\alpha,x}}$ can be determined as 
	\begin{equation}
		{{\mu }_{k,x}}=\frac{{{2}^{k}}-2}{{{2}^{k}}-1}+\frac{C_{l}^{k}}{\left( {{2}^{k}}-1 \right)C_{n}^{k}}\le 1,
	\end{equation}
	and the variance is 
	\begin{equation}
		\sigma _{k,x}^{2}={{\left( 1-{{\mu }_{k,x}} \right)}^{2}}{{\mu }_{k,x}}+\frac{\mu _{k,x}^{2}\left( C_{n}^{k}-C_{l}^{k} \right)}{\left( {{2}^{k}}-1 \right)C_{n}^{k}}\le \frac{1}{{{2}^{k}}-1},
	\end{equation}
	where $l=n-{{d}_{H}}(x,t)$ and $d_H(x,t)$ is the Hamming distance between binary strings $x$ and $t$. For the eigenvalue of $H_C$, ${\mathcal{E}}_{k,x} = \sum_\alpha E_{\alpha, x}$. It is notable that each $\alpha$ is independently selected from the same clause set. Consequently, $E_{\alpha, x}$ forms a sequence of independent and identically distributed (i.i.d.) random variables. According to the central limit theorem, ${\mathcal{E}}_{k,x}/m$ approximately follows the normal distribution such that
	\begin{equation}\label{eq_normal_distribution}
		\sqrt{m}\left( \frac{1}{m}{\mathcal{E}}_{k,x}-{{\mu }_{k,x}} \right)\sim N(0,\sigma _{k,x}^{2}).
	\end{equation}
	
	Evidently, the eigenvalue magnitude of $H_C$ is on the order of ${\mathcal O}(m)$, while for the mix Hamiltonian $H_B=\sum_j \sigma_x^{(j)}$, the eigenvalue magnitude is on the order of ${\mathcal O}(n)$. In the context of the $k$-SAT problem, the number of clauses $m$ can reach ${\mathcal O}(n^k)$ at most. This disparity results in a substantial energy difference between $H_C$ and $H_B$. Although this difference can be mitigated by adjusting ${\bm \gamma}$ and ${\bm\beta}$, the disparate magnitudes of parameters result in unnecessary difficulty during parameter adjusting. To address this, a preprocessing step can be employed to (approximately) normalize these two Hamiltonians, thereby reducing the maximal energy difference (the gap between the maximal and minimal energy) to 1.
	
\begin{figure}[!t]
		\centering
		\footnotesize
		{\includegraphics[width=0.9\linewidth]{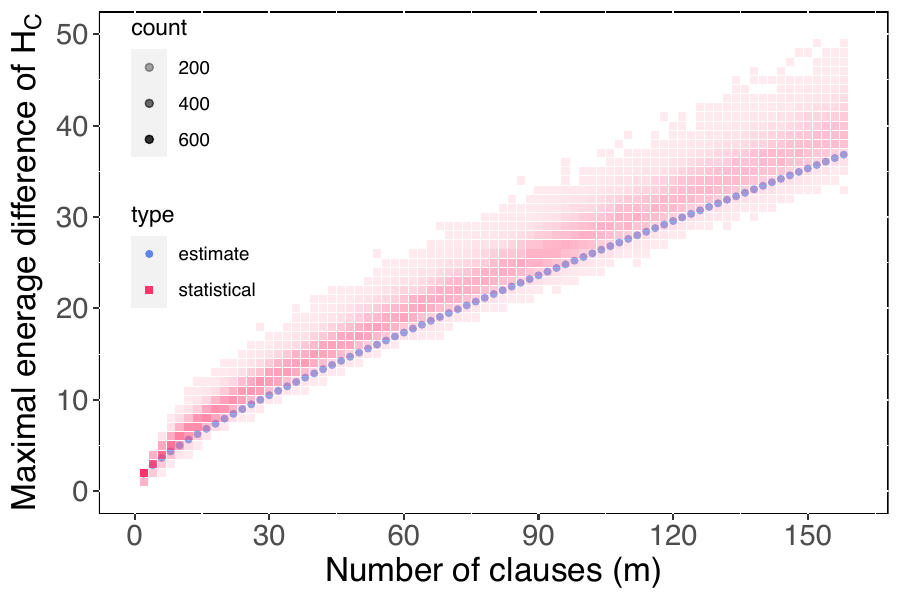}}
		\caption{The comparative results of $G_0$ between the estimation $G_E$ and the statistical results obtained from 1000 random instances in $F_s(20, m, 3)$ with varying values of $m$. The blue round points represent the estimation, while the red square points depict the statistical values, with different shades indicating the frequency of occurrence. The estimation serves as an approximate lower bound and closely aligns with the darkest square points in the figure. }
		\label{fig_exp0ge}
	\end{figure}

	The maximal energy difference of $H_B$ is determined as $2n$, while $H_C$ poses a challenge due to the exponential length of Hamiltonian in matrix form. However, leveraging the analysis of the problem Hamiltonian of random $k$-SAT, an estimate for $G_0$ can be derived. Referring to Eq.~(\ref{eq_normal_distribution}), it is anticipated that the majority of eigenvalues ${{\mathcal{E}}_{k,x}}$ should satisfy 
	\begin{equation}\label{eq_rangeOfE}
		{{\mu }_{k,x}}-\frac{{{c}_{0}}}{\sqrt{m({{2}^{k}}-1)}}\le \frac{1}{m}{{\mathcal{E}}_{k,x}}\le {{\mu }_{k,x}}+\frac{{{c}_{0}}}{\sqrt{m({{2}^{k}}-1)}},
	\end{equation}
	where ${{c}_{0}}$ is a constant not less than 3. Noting that $\max \{{{u}_{k,x}}\}=1$ and $ {{\mu }_{k,x}}+{{{c}_{0}}}/{\sqrt{m({{2}^{k}}-1)}} $ cannot exceed 1, an estimation of lower bound of $G_0$ can be obtained by the maximal difference of the majority of eigenvalues as 
	\begin{equation}
		\frac{1}{m}{{G}_{0}} \approx 1 -\min \left\{ {{\mu }_{k,x}}-\frac{{{c}_{0}}}{\sqrt{m({{2}^{k}}-1)}} \right\}.
	\end{equation}
	Therefore, the estimation $G_E$ can be determined as 
	\begin{equation}\label{eq_eGE}
		\frac{1}{m}{{G}_{E}} = \frac{1}{{{2}^{k}}-1}+\frac{{{c}_{0}}}{\sqrt{m({{2}^{k}}-1)}},
	\end{equation}
	This estimation, grounded in $F_f(n,m,k)$, can also be applied to $F_s(n,m,k)$ due to the similarity between these two models. An experimental comparison between this estimation $G_E$ and the actual $G_0$ of $F_s(n,m,k)$ is conducted, with the results presented in Figure \ref{fig_exp0ge}. 
	The simulation outcome demonstrates the effectiveness of this estimation for $F_s(n,m,k)$, thereby normalizing the Hamiltonians as 
	\begin{equation}\label{eq_norm_H}
		\bar{H}_B = \frac{H_B}{2n}, \bar{H}_C=\frac{H_C}{G_E}.
	\end{equation}
	
	\section{Parameter setting method}\label{sec_para_set}
	
	\subsection{QAA-inspired parameter initialization}\label{sec_para_init}
	
	Due to the similarity between the Trotterization of QAA and QAOA, a natural consideration is to initialize the parameters by a linearly varying $H(t)$, akin to the TQA initialization \cite{Sack2021}. Combined with the normalization of the Hamiltonian in Eq.~(\ref{eq_norm_H}), the parameterized system Hamiltonian should take the form of
	\begin{equation}\label{eq_paraform_two}
		\bar{H}(t)=\frac{t}{T}f_\gamma\bar{H}_C+\frac{T-t}{T}f_\beta\bar{H}_B
	\end{equation}
	with considerations for the degrees of freedom. Here, $f_\gamma$ and $f_\beta$ describe the intensities of the problem Hamiltonian and mix Hamiltonian in the evolution, respectively. After normalization, the energy of $\bar H_B$ and $\bar H_C$ should be in the same order of magnitude. Heuristically, the magnitudes of $f_\gamma$ and $f_\beta$ should also be similar. Therefore, by treating the intensities of $f_\gamma$ and $f_\beta$ as a single parameter $\rho$, the evolution of QAOA with linear-varying parameters can be viewed as the Trotterization of the evolution of the Hamiltonian 
	\begin{equation}\label{eq_H_theta_t}
		\bar{H}(\theta,t)=\frac{t}{T}\sin{\theta}\bar{H}_C+\frac{(T-t)}{T}\cos{\theta}\bar{H}_B, 
	\end{equation}
	with $T=2p\rho\pi$ and a discretization interval $\Delta t = 2\rho\pi$. In the context of QAOA with depth $p$, the parameters $(\gamma, \beta)$ can be expressed as 
	\begin{equation}
		\gamma_d=\frac{2d\pi}{(p+1)}\rho\sin{\theta}, \beta_d=\frac{2(p-d+1)\pi}{(p+1)}\rho\cos{\theta},
	\end{equation}
	where $1\le d \le p$. The choice of $p+1$ is made to avoid boundary situations where $\gamma_d$ or $\beta_d$ become 0 or $2\pi$.
	
	The initialization of the $2p$ parameters of QAOA is simplified by setting the values for $\theta$ and $\rho$. $\theta$ parameterizes the relative intensity ratio between $\bar{H}_C$ and $\bar{H}_B$. Due to the normalization of the Hamiltonian, it is natural to set $\theta = \pi/4$. The parameter $\rho$ describes the discretization width of QAA and also the evolution time of each layer of QAOA. Although there is no prior knowledge about the expected value of $\rho$, it has a natural initialization without drawbacks, specifically $\rho = \sqrt{2}$. The rationale behind this choice is as follows: In QAOA, the only problem-dependent operator is $H_C$, that is, the evolution $e^{-i \gamma H_C}$, which essentially induces phase shifts $e^{-i \gamma C(x)}$ on every computational basis $\left| x \right>$. Consequently, the target is distinguished by the phase shifts. To best distinguish the target $t$, $e^{-i \gamma C(t)}$ should be different from the others. Due to the periodicity of the phase, a fundamental consideration is to restrict the range of $\gamma C(x)$ to $\left[0, 2\pi\right)$. With the normalized $\bar{H}_C$, $f_\gamma$ can be initialized as 1, and $\rho = \sqrt{2}$.
	
	\begin{figure}[!t]
		\centering
		\footnotesize
		{\includegraphics[width=0.9\linewidth]{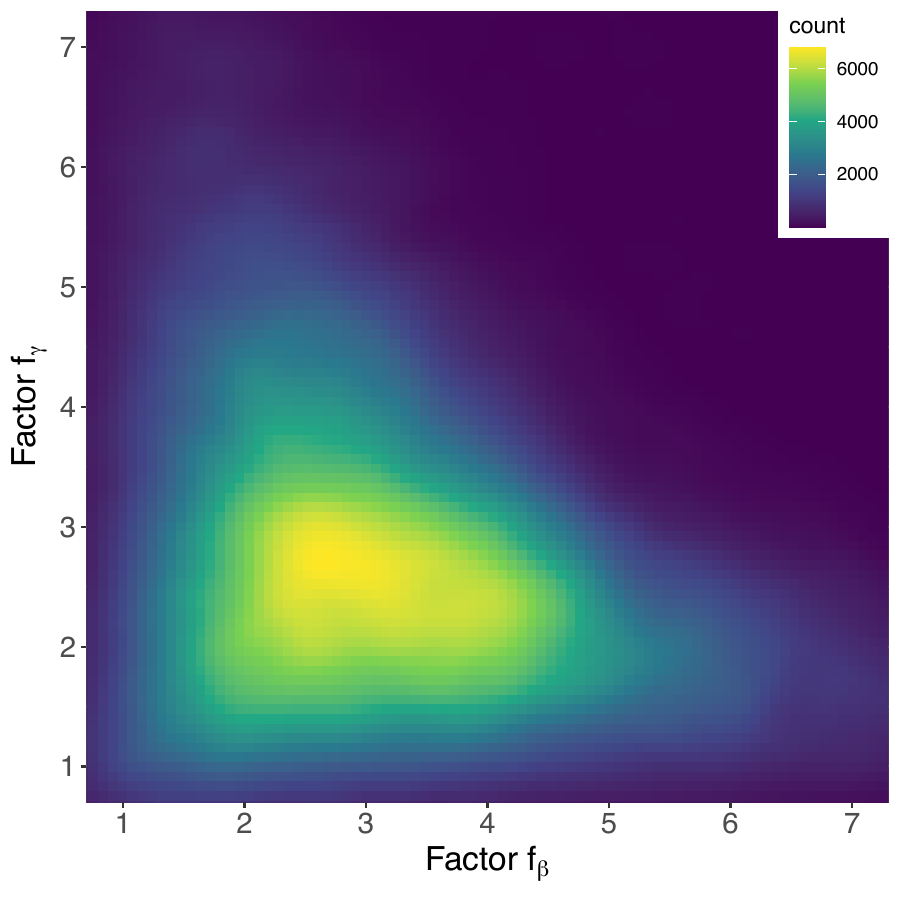}}
		\caption{\small The distribution of the count when obtaining the target state upon measurement. For each pair of $(f_\gamma, f_\beta)$, the final state is measured 10,000 times. This simulation is conducted on a random instance in $F_s(12,m,3)$. To emphasize the effect of normalization, $H_C$ is normalized by its real $G_0$. Furthermore, for application on NISQ devices, the depth $p$ of QAOA is fixed as $n$.} 
		\label{fig_exp4ng}
	\end{figure}
	
	A simulation has been conducted to validate the proposed initialization. For a random instance in $F_s(12,m,3)$, Figure \ref{fig_exp4ng} depicts the probability of the target state $\left| t \right>$ across various values of $f_\gamma$ and $f_\beta$. The simulation results reveal that the probability increases most rapidly around the axis $f_\gamma = f_\beta$ and soon reaches the optimum. Notably, in the vicinity of $( f_\gamma = 1, f_\beta=1)$, a pronounced gradient is observed toward the optimum. Deviation from this axis results in a diminished gradient and an increased distance from the optimum. Additionally, when $(f_\gamma, f_\beta)$ crosses a certain threshold, a notable decrease in probability occurs, with gradient nearly disappearing. This observation underscores the effectiveness of Hamiltonian normalization in setting the parameters.
	
	Notably, a larger $\rho$ implies more evolution time within the specific interval of evolution. Given the generally insufficient depth in QAOA, a reasonably larger $\rho$ corresponds to extended evolution time and tends to yield better performance. Consequently, as $n$ increases, while the optimal $\theta^*$ persists around $\pi/4$, the optimal $\rho^*$ should gradually increase with $n$, as illustrated in Figure~\ref{fig_exp4ng}. In this context,  the statistical optimum $(\bar{\theta^*}, \bar{\rho^*})$ for general instance can be approximately estimated from this distribution. 
	
	\begin{figure}[!t]
		\centering
		\footnotesize
		{\includegraphics[width=0.9\linewidth]{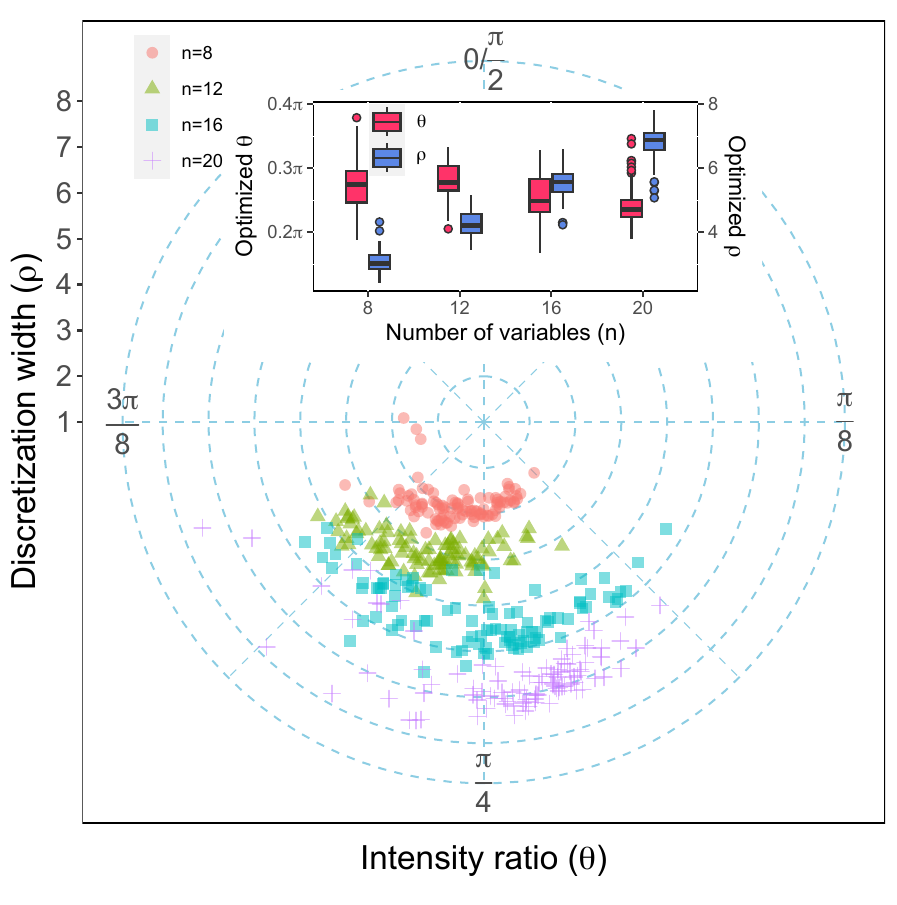}}
		\caption{\small The distribution of optimal $(\theta^*, \rho^*)$ for 100 random instances of $F_s(n,m,3)$ presented in a polar coordinate system. Different values of $n$, specifically $n=8,12,16,20$, are distinguished by different colors and shapes: red circles, green triangles, blue squares, and purple crosses, respectively. In the inner figure, the optimal $\theta^*$ generally centers around $\pi/4$, while $\rho^*$ exhibits a linear increase with $n$. Three idealized conditions are considered: first, the problem Hamiltonian $H_C$ is normalized by its accurate $G_0$; second, the probability of the target state is set as the optimization objective to maximize the fidelity; third, an exhaustive search is employed to identify the optimal parameters.}
		\label{fig_exp4id}
	\end{figure}

	The TQA initialization seeks to identify the statistical optimum $(\bar{\theta^*}, \bar{\rho^*})$ in the pre-computation stage. This consideration holds merit, particularly when adopting our normalized Hamiltonian, as it contributes to a substantial reduction in pre-computation costs owing to the direct guidance regarding the optimal region. However, for situations where optimal parameters $(\theta^*, \rho^*)$ significantly deviate from the estimates $(\bar{\theta^*}, \bar{\rho^*})$, initialization with a larger gradient is preferable, other than the overfitted parameters.  
	Consequently, the QAA-inspired parameter initialization method is formulated as presented in Algorithm \ref{alg1}. Moreover, this initialization method can be adapted into a QAA-inspired parameter setting method by optimizing $\left< H_C \right>$ through adjustments in $\theta$ and $\rho$. This modification is suitable for scenarios with strictly limited optimization costs.

	\subsection{Alternative parameter space based on continuity of adiabatic passage}\label{sec_para_space}
	
	The QAA-inspired initialization method enables the setting of $2p$ parameters without incurring any optimization cost or pre-computation, but it cannot fully exploit the computational potential of the circuit. 
	The inefficiency of QAA in utilizing the quantum circuit can be demonstrated by a simple modification, namely, segmented QAA. In segmented QAA, the entire evolution is divided into several segments as ${H(s_0), H(s_1),...,H(s_p)}$, and each segment $\left(H(s_{d-1}), H(s_{d})\right]$ applies QAA with a system Hamiltonian presented in Eq.~(\ref{eq_H_theta_t}) independently. In this approach, the minimal gap of each segment is only determined by $g(s)$ with $s \in \left(s_{d-1}, s_{d}\right]$, rather than the global minimum $g_0$. As not every segment has a minimal gap as $g_0$, the required evolution time in each segment is reduced. Consequently, the overall required time is decreased, leading to a reduction in the depth of the circuit.
	
		\begin{algorithm}[H]
		\caption{QAA-inspired initialization for QAOA}
		\small
		\label{alg1}
		\begin{algorithmic}[0]
			\Require~~\\ 
			Random model of problem $F$, QAOA's depth $p$;
			\Ensure~~\\ 
			Initial parameters $(\bm{\gamma}, \bm{\beta})$, normalized $\bar{H}_C$ and $\bar{H}_B$;
		\end{algorithmic}
		\begin{algorithmic}[1]
			\State Estimate the maximal energy difference $G_E$ of $H_C$ based on $F$, and set $\bar{H}_C \leftarrow H_C/G_E$ and $\bar{H}_B \leftarrow H_B/2n$; \Comment{Normalize Hamiltonians.}
			\State $\gamma_d \leftarrow \frac{2d\pi}{p+1}$, $\beta_d \leftarrow \frac{2(p+1-d)\pi}{p+1}$, and the evolution is $\prod_{d=1}^{p} e^{-i\beta_d \bar{H}_B}e^{-i\gamma_d \bar{H}_C}$.  \Comment{Set initial parameters.}
		\end{algorithmic}
	\end{algorithm}
	
	If the number of segments is large, each segment can be made small enough. For a small segment $\left(s_{d}- \Delta s, s_{d} \right]$, the minimal required evolution time should satisfy $T_d \sim \varepsilon(s_d)g^{-2}(s_d)$. As the segment size $\Delta s \to 0$, the required time for the optimal evolution in the entire interval $\left[0, 1\right]$ should satisfy \begin{equation}\label{eq_evolution_time}
		T \sim \int_{0}^{1}{\varepsilon(s)g^{-2}(s){\rm d}s}.
	\end{equation}
	To characterize this optimal evolution, the concept of adiabatic passage is introduced. According to the adiabatic theorem outlined in Section \ref{subsec_QAC}, $H(s)$ can be fixed to a certain extent for a specific problem. However, the mapping from $s$ to $t$ can vary, resulting in different possible system Hamiltonians $H(t)$. Regardless, $H(t)$ always adheres to the form
	\begin{equation}\label{eq_Ht_general}
		H(t)=f_C(t)H_C + f_B(t)H_B,
	\end{equation}
	where $f_C(0) = 0$, $f_B(0) > 0$, $f_C(T) > 0$, $f_B(T) = 0$, and $0 \le t \le T$. 
	As $t$ increases, $H(t)$ gradually transitions from $H_B$ to $H_C$, and the pair $(f_C(t), f_B(t))$ completely describes the precise evolution process, denoted as $\bm{f}(t)$, termed adiabatic passage (AP) in this paper. The optimal adiabatic passage achieves the minimal evolution time $T^*$ while ensuring sufficient fidelity. 
	Importantly, the optimal adiabatic passage $\bm{f}^*(t)$ is always continuous with respect to $t$. A straightforward proof can be offered by counter-evidence: if the optimal adiabatic passage are not continuous, a better adiabatic passage can be found by linear interpolation, resulting in a reduction in the evolution time by a smaller $\frac{{\rm d}H(t)}{{\rm d}t}$ in the discontinuous intervals.
	
	To simulate the optimal evolution, the optimal adiabatic passage is discretized into $p$ intervals. 
	For the $d$-th interval $\left(t_{d-1}, t_d\right]$, the evolution is approximated as 
	\begin{equation}\label{eq_AP_para}
		e^{iH(t_d)\Delta t} \approx e^{if^*_C(t_d)\Delta t_d H_C}e^{if^*_B(t_d)\Delta t_d H_B}, 
	\end{equation}
	where $\Delta t_d = t_d - t_{d-1}$. QAOA parameterizes the unknown time in the evolution with parameters $\gamma_d=f^*_C(t_d)\Delta t_d$ and $\beta_d=f^*_B(t_d)\Delta t_d$, aiming to approximate the optimal adiabatic passage $\bm{f}^*(t)$ through adjustments in parameters. If there is no information about $\bm{f}^*(t)$, this straightforward parameterization is optimal. However, $\bm{f}^*(t)$ is known to be continuous, providing an opportunity for a more refined parameterization of $\bm{f}(t)$.
	
	In the evolution described by Eq.~(\ref{eq_AP_para}), three explicit factors are involved: $\Delta t_d$, $f^*_C(t_d)$, and $f^*_B(t_d)$. The relationship among these three factors is determined by the optimality of $f^*_C(t)$ and $f^*_B(t)$. Consequently, according to the adiabatic approximation, the evolution time $\Delta t$ follows 
	\begin{equation}\label{eq_factor_relation}
		\Delta t_d\sim \varepsilon(t_d)g^{-2}(t_d).
	\end{equation}
	Based on Eq.~(\ref{eq_T_varepsilon}), the main factors influencing $\varepsilon(t_d)$ are $\frac{{\rm d}H(t)}{{\rm d}t}$. Therefore, it is $(\Delta f^*_C(t_d), \Delta f^*_B(t_d))$ that directly affects $\Delta t_d$, rather than $(f^*_C(t_d), f^*_B(t_d))$. In reality, four internally related factors are involved in the evolution, namely, $\Delta t_d$, $g(t_d)$, $\Delta f^*_C(t_d)$, and $\Delta f^*_B(t_d)$. Here for convenience, denote $\Delta f^*_C(t_d) = f^*_C(t_d) - f^*_C(t_{d-1})$ and $\Delta f^*_B(t_d)=f^*_B(t_{d-1})-f^*_B(t_d)$.

	Because the gap $g(t)$ is an inherent property of the Hamiltonian that cannot be adjusted by parameters, it can be regarded as an independent variable. The other three factors depend on $g(t_d)$ and correspondingly exhibit different levels of continuity. In fact, discrete variables have no continuity in the strict sense, and better continuity here refers to a smaller difference in the corresponding factors between adjacent intervals. Based on Eq.~(\ref{eq_factor_relation}), $\Delta f^*_C(t_d)$ and $\Delta f^*_B(t_d)$ should be quadratically positively correlated with $g(t_d)$, while $\Delta t_d$ is quadratically inversely related to $g(t_d).$ Besides, these factors are also restricted by the Trotterization constraint, thus the magnitude of the factors cannot be too large. Specifically, when $g(t_d)$ becomes small, $\Delta f^*_C(t_d)$ and $\Delta f^*_B(t_d)$ mainly reduce, rather than quadratically increasing $\Delta t_d$.

	The continuity of parameters is crucial in parameter setting; therefore, the parameterization of the adiabatic passage should maintain the best continuity of these factors. Obviously, $\Delta f^*_C(t_d)$ exhibits better continuity compared to the summation $f^*_C(t_d) = \sum_{j=1}^d \Delta f^*_C(t_j)$. Besides, according to the analysis of the normalized Hamiltonian in Section \ref{sec_preproc} and \ref{sec_para_init}, $\Delta f^*_C(t_d)$ and $\Delta f^*_C(t_d)$ should exhibit similar properties to $f_\gamma$ and $f_\beta$. Consequently, $(\Delta f^*_C(t_d), \Delta f^*_C(t_d))$ can be mapped as $(\rho_d\sin{\theta_d}, \rho_d\cos{\theta_d})$. A straightforward consideration is to directly parameterize these three dependent factors and its mapping to $({\bm \gamma}, {\bm \beta})$ can be expressed as
	\begin{equation}\label{eq_para3}
		\gamma_d=\sum_{j=1}^d \rho_j\sin{\theta_j}\Delta t_d,
		\beta_d=\sum_{j=d}^p \rho_j\cos{\theta_j}\Delta t_d. 
	\end{equation}
	Here, $\theta_d$ represents the intensity ratio between the increments of $H_C$ and $H_B$ in the $d$-th interval and has an approximate expectation of $\pi/4$ if the Hamiltonian is properly normalized. It certainly exhibits good continuity. On the other hand, $\rho_d$ describes the intensity of ${\rm d}H(t)$ and is positively correlated with $g(t_d)$. Comparing it to the individual components $\Delta f^*_C(t_d)$ or $\Delta f^*_C(t_d)$ alone, $\rho_d$ exhibits better continuity. 
	
	This ternary parameterization provides good continuity and interpretability in parameters. 
	However, the presence of $3p$ parameters introduces higher optimization costs. 
	To reduce the number of parameters, an approximation can be made by extracting the increasing (or decreasing) factor in the summation alone, specifically for $H_C$ as an example, denoted as
	\begin{equation}
		\kappa_d = \frac{1}{p\rho_d\sin{\theta_d}}\sum_{j=1}^{d}\rho_j\sin{\theta_j}. 
	\end{equation} 
	The term $\rho_j\sin{\theta_j}$ for each layer can be approximately regarded as i.i.d. random variables. As $d$ increases, a linear increase would gradually dominate, and this factor should take a form as $\kappa_d =\frac{d}{p}+\Delta\kappa_d$.
	
	In Eq.~(\ref{eq_para3}), $\gamma_d$ can be rewritten as $p \kappa_d \rho_d \Delta t_d \sin{\theta_d}$. Since $\rho_d$ and $\Delta t_d$ exhibit opposite correlations with $g(t_d)$, $\rho_d\Delta t_d$ demonstrates better continuity. On the other hand, $\sum_{d=1}^p \Delta t_d = T$, so $ \Delta t_d $ is approximately inversely proportional to $p$. Consequently, $p \rho_d \Delta t_d$ can be combined into a single parameter $\tau_d$ with improved continuity. Regarding $\kappa_d$, the lower-order term $\Delta \kappa$ can be divided into intensity effects and angle effects, merging into $\tau_d$ and $\theta_d$, respectively. As a result, the parameter space is reduced to $({\bm\theta, \bm\tau})$, and its mapping with the original $({\bm \gamma}, {\bm \beta})$ is
	\begin{equation}
		\gamma_d=\frac{d}{p+1}\tau_d\sin{\theta_d}, \beta_d=\frac{p+1-d}{p+1}\tau_d\cos{\theta_d}, 
	\end{equation}
	where the choice of $p+1$ is made to avoid boundary situations. 
	
	\subsection{Adiabatic-passage-based parameter setting method}\label{sec_para_AP}
	
	By employing a specific initialization, the resulting parameters can be conceptualized as an estimate of the optimal adiabatic passage $\bm{f}^*(t)$. The parameter setting process aims to minimize the distance from the estimation $\bm{f}(t)$ to the optimum $\bm{f}^*(t)$ through parameter adjustment, specifically, minimizing $\left\| \Delta \bm{f}(t)\right\|$, where $\Delta \bm{f}(t) = \bm{f}_{C}(t) - \bm{f}^*(t)$. Given the continuity of the function, the norm $\left\| f(x) \right\| = \max_x\{ \left| f(x) \right| \}$ is employed. According to the method in Section \ref{sec_para_init}, a substantial gradient can be anticipated around the initialization, guiding the optimization in parameter adjustment. 
	Consequently, apart from the precision $\epsilon$, the primary factor influencing the optimization cost is the distance $\left\| \Delta \bm{f}(t)\right\|$. When $\left\| \Delta \bm{f}(t)\right\|$ is constant, constant gradient calculations are required, with each gradient calculation necessitating $\mathcal{O}(p)$ circuit runs. Given the distribution of optimal $(\theta^*, \rho^*)$ depicted in Figure \ref{fig_exp4id}, it is anticipated that $\left| \Delta \bm{f}(t)\right|$ is generally bounded with the normalized Hamiltonian. If further normalizing the magnitude of parameters to be $\mathcal{O}(n)$, constant gradient calculations can be achieved. 
	
	As a result, the optimization cost is reduced to approximately linear with respect to $p$ in the idealized situation. Actually, for this type of parameter adjusting process, a kind of ``synchronous adjustment'' exhibits better performance. The primary objective of the parameterized adiabatic passage is to utilize $p$ ``samplings'' to simulate the optimal adiabatic passage. Therefore, it is more practical to commence the process with a small number of sampling points to adjust the main components. and then gradually increase the number to $p$ to better align with the details of optimal adiabatic passage, akin to the FOURIER strategy introduced in \cite{PRX2020}.
	
	Specifically, FOURIER introduces an indicator $q$ to represent the current number of sampling points. It decomposes $\bm{f}(t)$ into $q$ components using the cos/sin Fourier transform. With a linear increase in $q$, the optimal parameters from the previous iteration can be utilized as the initialization, resulting in the distance $\left\| \Delta \bm{f}_{d}(t)\right\|$ for the $d$-th component being approximately reduced to less than $\left\| \Delta \bm{f}(t)\right\|/q$ on average. This reduction contributes to a decrease in optimization cost in each iteration. However, with $\mathcal{O}(n)$ iterations in the outer loop, the optimization cost persists superlinear. 
	
	It is noteworthy that the approximate adiabatic passage with any number of sampling points can serve as an initialization for another approximation with more sampling points by interpolation. Consequently, we refine this approach by doubling the sampling points for each iteration other than linear increasing, aiming to mitigate the optimization cost from the iterations of outer loop. In the $l$-th iteration, $2^l$ sampling points are employed, and the initial parameters are obtained through interpolation of previous $2^{l-1}$ parameters. Assuming sufficient continuity in the parameter space, the distance $\left\| \Delta f_j(t)\right\|$ of the $j$-th interval, where $1\le j \le 2^l$, can be approximately reduced to a level of $\left\| \Delta f(t)\right\|/2^l$, resulting in only logarithmic iterations and potentially an overall cost of $\mathcal{O}(\log(p))$. The corresponding parameter setting method is presented in Algorithm \ref{alg2}.
	
	\begin{algorithm}[H]
		\caption{AP-based parameter setting for QAOA}
		\label{alg2}
		\small
		\begin{algorithmic}[0]
			\Require ~~\\ 
			Normalized $\bar{H}_C$ and $\bar{H}_B$, 
			routine $Interp(\bm{v},L)$ that interpolates vector $\bm{v}$ to a length of $L$;
			\Ensure ~~\\ 
			The quasi-optimal  parameters $(\bm{\theta}, \bm{\tau} )$;
		\end{algorithmic}
		\begin{algorithmic}[1]
			\State Initial $\theta_0 = \pi/4, \tau_0 = \sqrt{2}$, and optimize $\theta_0, \tau_0$ by the QAA-inspired setting to $\theta^*_0$ and $\tau^*_0$;
			\State $\bar{H}_C \leftarrow \sqrt{2}\tau^*_0\sin\theta^*_0\bar{H}_C$,  $\bar{H}_B \leftarrow \sqrt{2}\tau^*_0\cos\theta^*_0\bar{H}_B/ $, $T_u \leftarrow \left\lfloor \log_2{p} \right\rfloor$, and initialize $\bm{\theta}\leftarrow(\pi/4)$, $\bm{\tau}\leftarrow(1)$;
			\State If $T_u \le 0$, finish; otherwise, $T_u \leftarrow T_u-1$, $\bm{\theta} \leftarrow Interp(\bm{\theta}, \left\lceil {p/2^{T_u}} \right\rceil)$, $\bm{\tau} \leftarrow Interp(\bm{\tau}, \left\lceil {p/2^{T_u}} \right\rceil)$;
			\State $\bm{\theta'} \leftarrow Interp(\bm{\theta}, p)$, $\bm{\tau'} \leftarrow Interp(\bm{\tau}, p)$, adjust $\bm{\theta}, \bm{\tau}$ to optimize $\left< \bm{\theta'}, \bm{\tau'} \left | H_C \right|  \bm{\theta'}, \bm{\tau'}  \right>$, then jump to Step 3.
		\end{algorithmic}
	\end{algorithm}

	In Require, this algorithm inherits the normalized Hamiltonian in Algorithm \ref{alg1}. Besides, the interpolation routine is flexible, and in this paper, cubic spline interpolation is employed. The explanation of each step is as follows. In Step1, the optimal parameters for ${\bm f}^*(t)$ with one sampling are obtained. Due to the refined parameter space, the magnitude of parameters with any number of sampling points remains consistent. Consequently, Step 2 normalizes the parameter magnitude by introducing the optimized $\rho^*_0,\tau^*_0$ as coefficients into the Hamiltonian. Step 3 and Step 4 gradually double the number of sampling points until reaching $p$. For each iteration, the parameters are optimized and utilized as initialization for the next iteration.

	\section{Comparison and performance analysis}\label{sec_comp_pref} 
	
	This section conducts a comparative analysis between the methodologies presented in this paper and those from previous works, specifically the TQA initialization in \cite{Sack2021}, along with the INTERP heuristic and FOURIER heuristic proposed in \cite{PRX2020}. Furthermore, an analysis is dedicated to the performance of the AP-based method, elucidating the observed logarithmic increase in optimization cost.	
	
	Some details of the simulation are outlined as follows. 
	Firstly, the depth $p$ maintains fixed at $n$ for the application on NISQ devices.
	Secondly, the performance of QAOA is evaluated by the probability of the target state, in accordance with the problem reduction in Section \ref{subsec_k_SAT}. Regarding the parameter setting method, the performance is assessed based on the optimization cost, specifically, the required circuit runs to achieve a quasi-optimum of parameters. Noteworthily, this is distinct from the number of epochs, which entails the calculation of the full gradient and consequently involves $\mathcal{O}(p)$ circuit runs. In the simulation, the required times to calculate the expectation $\left< H_C \right>$ are presented. 
	Thirdly, the number of clauses $m$ is given by an experimental value $m_n^*$, where the average number of interpretations is approximately 1.3. In the simulation, $m_n^* = \mu_n n$, with $\mu_n$ slightly increasing from 5.9 to 6.3 as $n$ ranges from 10 to 20. This is indicated as a challenging situation in \cite{Achioptas2006, Kirousis1998, Coja2016}.
	Finally, a gradient-based optimization routine (BFGS) is employed for classical optimization, corresponding to the approach in \cite{PRX2020} for the sake of comparison.
	
	\subsection{Comparison of performance under linear-varying parameters}
	
\begin{figure}[!t]
		\centering
		\footnotesize
		{\includegraphics[width=0.95\linewidth]{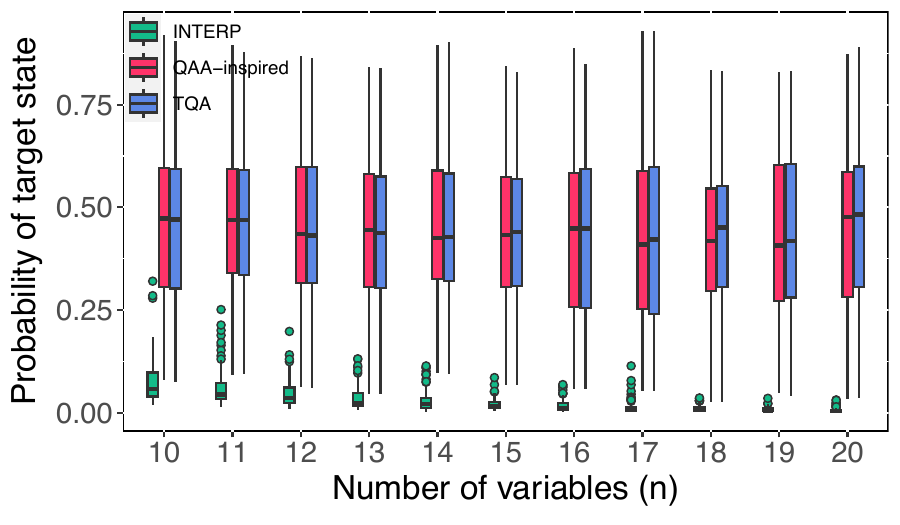}}
		\caption{The distribution of the target state probability for 100 random instances in $F_s(n,m_n^*,3)$ with $10 \le n \le 20$. The results of the INTERP heuristic, the QAA-inspired method and the TQA method are respectively represented in green on the left, red in the middle, and blue on the right. }
		\label{fig_exp_com2a}
	\end{figure}

\begin{figure}[!t]
		\centering
		\footnotesize
		{\includegraphics[width=0.95\linewidth]{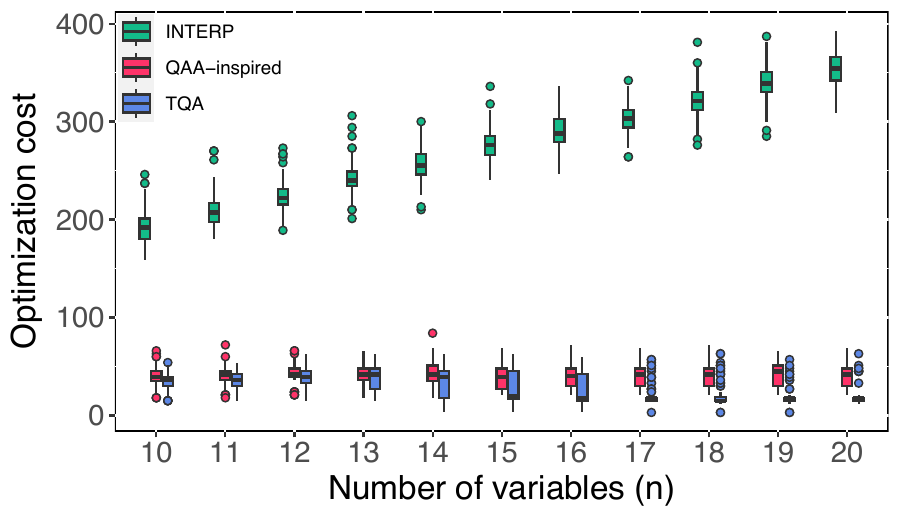}}
		\caption{The distribution of optimization costs for 100 random instances in $F_s(n,m_n^*,3)$ with $10 \le n \le 20$. The outcomes of the INTERP heuristic, the QAA-inspired method and the TQA method are respectively represented in green on the left, red in the middle, and blue on the right. } 
		\label{fig_exp_com2b}
	\end{figure}

	The QAA-inspired initialization can be adapted into a parameter setting method involving two parameters $(\theta, \rho)$, consequently resulting in linear-varying parameters $({\bm\gamma}, {\bm\beta})$. The TQA initialization has a similar transformation, with the only distinction lying in the initial $(\theta, \rho)$ to be the statistical optimum $(\bar{\theta}^*, \bar{\rho}^*)$ obtained by pre-computation, other than analytical values $(\pi/4, \sqrt2)$. The INTERP heuristic also employs linearly varying parameters, with the specific steps presented in Algorithm \ref{alg_INTERP}. Notably, the INTERP heuristic circumvents the discussion of parameters initialization by commencing with a small depth. Consequently, the initialization of QAOA with a larger depth is obtained by the optimized parameters of QAOA with a smaller depth. A comparative simulation is conducted for these methods on random instances in $F_s(n,m_n^*,3)$. The performance in terms of optimization cost and probability is illustrated in Figure \ref{fig_exp_com2a} and \ref{fig_exp_com2b}. 
	
	\begin{algorithm}[H]
		\caption{INTERP heuristic for QAOA}
		\label{alg_INTERP}
		\small
		\begin{algorithmic}[1]
			\State Randomly initialize $\gamma_0, \beta_0$ in range $\left[0, 2\pi\right]$, $q \leftarrow 1$;
			\State $\gamma_d \leftarrow \frac{d\gamma_0}{q+1}$, $\rho_d \leftarrow \frac{(q+1-d)\beta_0}{q+1}$, where $1 \le d \le q$; adjust $\gamma_0, \beta_0$ to optimize $\left< \bm{\gamma}, \bm{\beta} \left | H_C \right| \bm{\gamma}, \bm{\beta} \right>$;
			\State If $q \ge p$, terminate; otherwise, set $q\leftarrow q+1$ and jump to Step 2. 
		\end{algorithmic}
	\end{algorithm}	
	
	The QAA-inspired parameter setting method demonstrates superior performance compared to the INTERP heuristic. Additionally, the QAA-inspired method shows comparable performance with the TQA method without incurring any pre-computation. It is noteworthy that, in the absence of the normalized Hamiltonian, the pre-computation cost for TQA initialization would significantly increase with $n$ in the worst-case scenario.
	
	The high probability achieved by the QAA-inspired method is primarily attributed to the normalized Hamiltonian and the corresponding feasible initialization. Without this normalization, the optimal $\gamma_0$ for INTERP on 3-SAT diminishes with increasing $m$. Consequently, the performance of INTERP exhibits a high dependence on the initial parameter, specifically, whether $\gamma_0$ is initialized sufficiently small. When $\gamma_0$ is initialized with a random value in the range $[0,2\pi]$, the optimization process tends to get trapped in a suboptimal local minimum. 
	In terms of optimization cost, the INTERP method exhibits linear growth with $n$ due to the outer loop with $p$ iterations. Conversely, the QAA-inspired method, benefiting from the analytical initialization, involves optimization with two parameters without an outer loop, naturally resulting in a constant cost. Additionally, due to the optimum $(\theta^*, \rho^*)$ distributing around the statistical optimum $(\bar{\theta}^*, \bar{\rho}^*)$, the TQA method requires fewer circuit runs, although the improvement is marginal. 
	
	\subsection{Comparison of performance under quasi-optimal parameters}

	The AP-based method optimizes $2p$ parameters and exhibits similarities with the FOURIER heuristic, aiming to reduce optimization costs by capitalizing on parameter continuity. Specifically, FOURIER constrains the degree of freedom in the parameter space to a value of $q$ and then derives the complete parameters through Sin/Cos Fourier transform. Similarly, the AP-based method introduces an indicator $T_u$ to control the degree of freedom and obtains the full parameters through interpolation. Although both methods may terminate with parameter spaces of insufficient degrees of freedom, for the purpose of performance comparison, optimization is conducted with the full parameter space. The steps of the FOURIER heuristic in this context are presented in Algorithm \ref{alg_FOURIER}. 
	
\begin{figure}[!t]
		\centering
		\footnotesize
		{\includegraphics[width=0.95\linewidth]{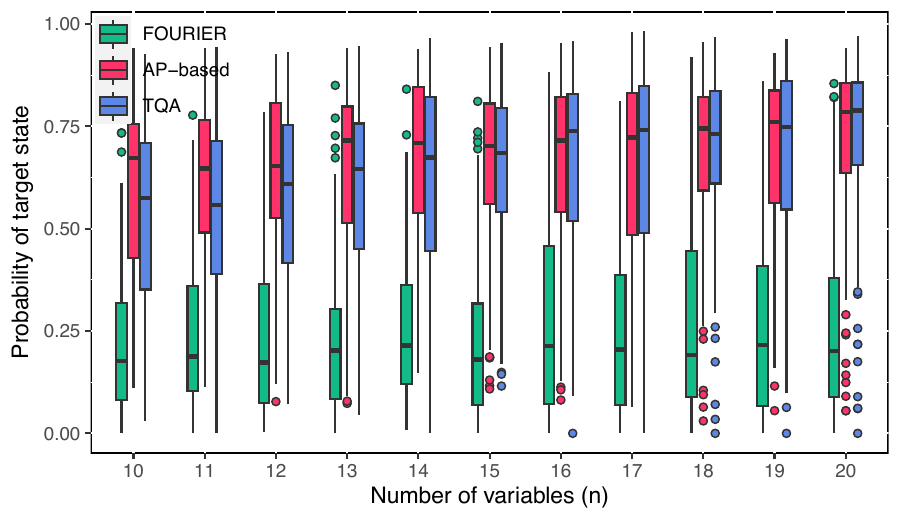}}
		\caption{This distribution of the target state probability for 100 random instances in $F_s(n,m_n^*,3)$ with $10 \le n \le 20$. The results of the FOURIER heuristic, the AP-based method and the TQA method are respectively represented in green on the left, red in the middle, and blue on the right.}
		\label{fig_exp_com1a}
	\end{figure}

\begin{figure}[!t]
		\centering
		\footnotesize
		{\includegraphics[width=1\linewidth]{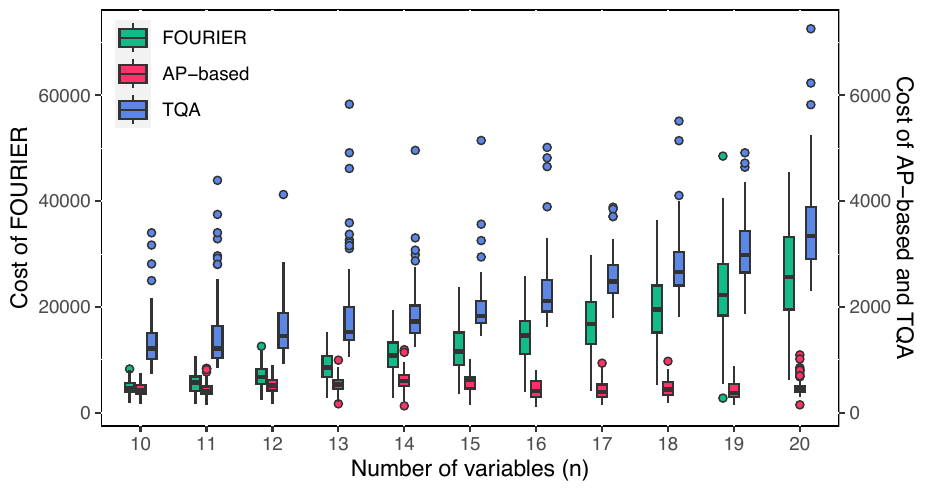}}
		\caption{ The distribution of optimization costs for 100 random instances in $F_s(n,m_n^*,3)$ with $10 \le n \le 20$. The results of the FOURIER heuristic, the AP-based method and the TQA method are respectively represented in green on the left, red in the middle, and blue on the right. Due to the difference in the amplitude of these results, a dual y-axis is utilized to provide a clearer representation of the cost.}
		\label{fig_exp_com1b}
	\end{figure}

	Noteworthily, the FOURIER heuristic is primarily designed for situations lacking proper initialization, leading to $\mathcal{O}(n)$ iterations in the outer loop. In contrast, the AP-based method inherits the advantage of the QAA-inspired method, initiating with a well-established QAOA initialization at the full depth. Furthermore, the AP-based method transforms the parameter space of $({\bm\gamma}, {\bm\beta})$ into another with enhanced continuity. The TQA method also achieves a good initialization through pre-computation and can also be utilized to optimize the full parameters by directly employing the initial parameters in the optimization process without additional strategy. A comparative simulation is conducted for these three methods on 100 random instances in $F_s(n,m_n^*,3)$, and the performance is presented in Figures \ref{fig_exp_com1a} and \ref{fig_exp_com1b}.
	
	\begin{algorithm}[H]
		\caption{FOURIER heuristic for QAOA}
		\label{alg_FOURIER}
		\small
		\begin{algorithmic}[1]
			\State Randomly initialize $\bm{u}=(u_1), \bm{v}=(v_1)$, $q \leftarrow 1$;
			\State Obtain $(\bm{\gamma}, \bm{\beta})$ of length $q$ by Sin/Cos transform $\gamma_j=\sum_{k=1}^q u_k \sin{((k-\frac{1}{2})(j-\frac{1}{2})\pi / q)}$, $\beta_j=\sum_{k=1}^q v_k \cos{((k-\frac{1}{2})(j-\frac{1}{2})\pi / q)}$; adjust $(\bm{u},\bm{v})$ to optimize $\left< \bm{\gamma}, \bm{\beta} \left | H_C \right| \bm{\gamma}, \bm{\beta} \right>$ using QAOA of a depth $q$;
			\State If $q \ge p$, terminate; otherwise, expand parameters $\bm{u}\leftarrow(u_1,...,u_{q},0)$, $\bm{v}\leftarrow(v_1,...,v_{q},0)$, and set $q\leftarrow q+1$. Jump to Step 2. 
		\end{algorithmic}
	\end{algorithm}

	
	Comparing with FOURIER, both the AP-based and TQA methods exhibits superior performance in terms of the success probability. The noteworthy success of FOURIER on max-cut problems in \cite{PRX2020} underscores the importance of Hamiltonian normalization and parameter initialization for problems like 3-SAT, where the intensity of the problem Hamiltonian experiences rapid growth. Additionally, in this paper, the parameters are initialized only once, resulting in poor performance for these heuristic strategies, which might get trapped in a local optimum due to the absence of a known feasible initialization.	
	
	In terms of optimization cost, FOURIER exhibits growth slightly faster than ${\mathcal O}(n^2)$, with an average of 4853.65 when $n=10$ and 26121.76 when $n=20$. This magnitude results from two factors: the linear increase in depth $q$, similar to INTERP, and the expanded degree of freedom of parameters from 1 to $q$. For the TQA method, due to the proper initialization obtained through pre-computation, the outer loop is eliminated, resulting in reduced optimization costs slightly larger than $\mathcal{O}(n)$, with an average of 1326.99 when $n=10$ and 3523.13 when $n=20$. Notably, the AP-based method further reduces the cost to a sublinear level. Further details about the cost will be discussed in Section \ref{subsec_perform}.
	
	\subsection{Performance analysis}\label{subsec_perform}

	The average optimization costs of the AP-based method within the range of $4 \le n \le 20$ are outlined in Table \ref{tab_count}. With the increase in the number of variables $n$, both the depth $p$ and the number of parameters $2p$ undergo augmentation. Correspondingly, the optimization cost experiences growth, demonstrating a sublinear increase in the overall trend, with values of 300.5 when $n=4$ and 546.3 when $n=20$. However, it is noteworthy that the cost does not exhibit a strict increase in simulation. Instead, this growth appears to be confined to specific intervals and exhibits fluctuations upon transitioning between them. 
	
	\begin{table}[!t] 
		\footnotesize
		\caption{The average optimization cost of AP-based method for 1000 random instances in $F_s(n,m_n^*, 3)$ with $4 \le n \le 20$. The range of $n$ is partitioned by the logarithmic indicator $\left\lfloor \log_2{p} \right\rfloor$, resulting in three intervals: $4 \le n \le 7$, $8 \le n \le 15$, and $16 \le n \le 20$, respectively. }
		\label{tab_count}
		\tabcolsep 2.75pt 
		\begin{tabular*}{0.44\textwidth}{ccccccccc}
			\toprule
			$n$ & [4 & 5 & 6 & 7] & [8& 9 & 10 & 11 \\\hline
			cost & $300.5$& $342.9$& $379.5$& $410.6$& $430.4$& $440.3$& $445.0$ & $446.4$\\
			\bottomrule
			12 & 13 & 14 & 15] & [16 & 17 & 18 & 19 & 20]\\\hline
			$532.2$& $544.3$& $586.0$& $612.2$& $524.9$& $535.7$& $573.5$& $531.7$& $546.3$\\
			\bottomrule
		\end{tabular*}
	\end{table}
	
	The intervals are demarcated by the logarithmic indicator $T_u$ with a value $l=\left\lfloor \log_2{p} \right\rfloor$, which controls the outer loop of AP-based method for increasing parameters. Precisely, the range of $n$ is divided into three intervals: $4 \le n \le 7$, $8 \le n \le 15$, and $16 \le n \le 20$. Within each interval of $n$, $l$ remains constant, while the number of parameters undergoes incremental changes. When $n=2^l$, this interpolation method is best suited, ensuring an accurate doubling of the parameter space with every iteration. With further increasing $n$, the efficiency diminishes due to the reduced continuity between parameters by interpolation, leading to an increase in cost. Thus, the most representative results are expected when $n=2^l$. For cases where $n=4,8,16$, the average costs are 300.5, 430.4, and 524.9, respectively, indicating an approximate linear relationship with $l$. This observation suggests the potential for the optimization cost to exhibit a logarithmic relationship with the depth $p$.
	
	Moreover, specifically considering the case of $n=16$, the optimization cost gradually rises with the decrease in the indicator $T_u$, leading to a corresponding doubling of the degree of freedom of the parameters for each iteration. The accumulated costs for each iteration are 40.7, 123.1, 294.7, 448.6, and 524.9, respectively. This cost exhibits an approximately linear increase concerning $l-T_u$. Notably, the last three results are nearly identical to the full cost when $n=4,8,16$. This observation reveals that the optimization cost of this method appears to have little correlation with $n$ or $p$ but rather with the degree of freedom of parameters. Consequently, to achieve the best efficiency of this method, the degree of freedom for parameters can be set as $2^l$ or $2^{l+1}$ in practical applications when $p \neq 2^l$. In such scenarios, the performance in terms of success probability would be nearly maintained due to the continuity of optimal adiabatic passage.  
	
\begin{figure}[!t]
		\centering
		\footnotesize
		{\includegraphics[width=1\linewidth]{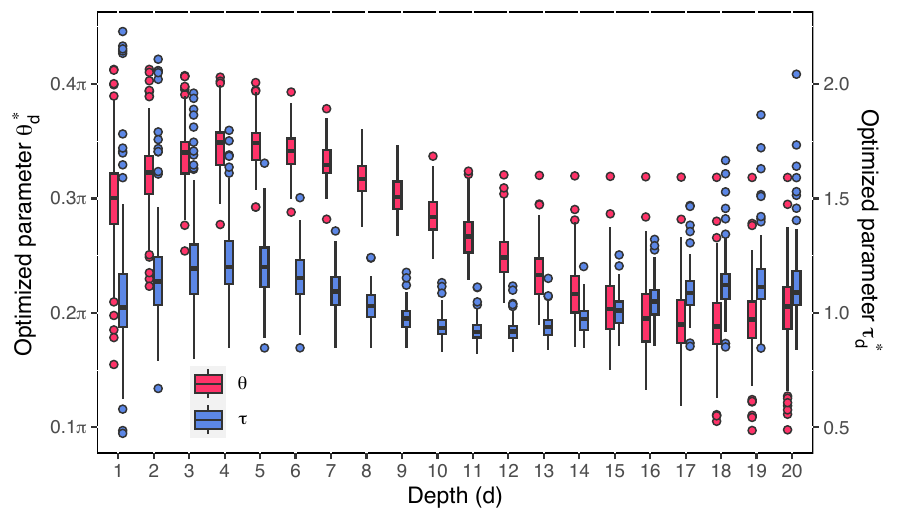}}
		\caption{ The distribution of the optimized parameters ${\bm \theta}^*$ and ${\bm \tau}^*$ for 100 random instances in $F_s(20,m_n^*,3)$, presented in red on the left and blue on the right, respectively. The optimized parameters general present superior continuity reminiscent of trigonometric functions, with even the outliers persisting very close to the adjacent parameters. }
		\label{fig_exp6para}
	\end{figure}

	The efficiency of the proposed method can be attributed to the continuity of optimal adiabatic passage, illustrated by the optimized parameters presented in Figure \ref{fig_exp6para}. These optimized parameters exhibit a smooth continuity reminiscent of trigonometric functions, with $\theta_d$ and $\rho_d$ centered around $\pi/4$ and $1$, respectively. In simulation, the impact of the variation in $g(s)$ is significantly reduced, leading to $\max_d\{\tau_d\}$ being no more than twice as much as $\min_d\{\tau_d\}$ . Conversely, for $g(s)$, the maximum generally differs by several times from the minimum. This outcome aligns with the analysis detailed in Section \ref{sec_para_set}, emphasizing the significance of Hamiltonian normalization and parameter continuity. Additionally, it reveals the potential suitability of the Sin/Cos Fourier transform in leveraging the continuity of parameters spaces $({\theta},{ \tau})$. The interpolation routine is employed in this paper primarily due to its simplicity.

	In practical implementation, the expectation is obtained through multiple repetitions of the circuit with the given parameters, typically involving 1000 circuit runs. The circuit complexity of a single circuit run depends on $H_C$ and $p$, specifically ${\mathcal O}(mp)$. Although it seems to suggest a large total cost, in reality, due to the preprocessing and the initialization, the circuit of QAOA exhibits commendable performance initially that requires only a few circuit runs, as revealed in Figure \ref{fig_exp_com2a}. Nonetheless, the parameter setting method remains crucial due to the problem reduction in Section \ref{subsec_k_SAT} when handling unsatisfiable instances. Given an acceptable time $T={\mathcal O}({\rm Poly}(n))$ to terminate the optimization, an efficient parameter setting method converges more rapidly to the quasi-optimum of parameters, resulting in a smaller error rate. Consequently, the algorithm also applies on general instances of 3-SAT.

	\section{Discussion}\label{sec_discussion}
	
	While the analyses and simulations in this paper are centered around 3-SAT problem, the proposed framework is universally applicable. For other combinatorial optimization problems, a random model can also be established. Building on this, the random Hamiltonian of this model can be constructed, where different techniques might be required, including problem reduction, model approximation, etc. Subsequently, by analyzing the statistical properties of the random Hamiltonian, it becomes possible to derive an estimate for the maximum energy difference of $H_C$. Consequently, the parameter setting method outlined in this paper remains applicable in this broader context.
	
	For combinatorial optimization problems related to graphs, such as max-cut and max-clique, natural random models like the Erd\H{o}s-R\'enyi model are available.  
	The distribution of eigenvalues for the given problem Hamiltonian can be established accordingly. In \cite{PRX2020}, max-cut on 3-regular graph is mainly discussed. It is noteworthy that the maximum eigenvalue of $H_C$ cannot exceed the total number of edges, i.e., $1.5n$, remaining in the same magnitude with that of $H_B$. Consequently, the INTERP and FOURIER heuristic demonstrate a high probability of reaching the target state. However, when regarding k-regular graph with a larger $k$ or random graphs $G(n, \rho)$, the normalization of Hamiltonian becomes crucial for achieving a good initialization. 
	Besides, for constrained combinatorial optimization like max-clique, the problem Hamiltonian varies with the specific design of goal function $C(x)$, but the distribution can still be derived.
	
	Certainly, there are combinatorial optimization problems that lack established popular random models. Nevertheless, in the domain of combinatorial optimization, a comprehensive set of problem instances always exists, providing the opportunity to design specific random models. Taking the set cover problem as an example, a random model can be constructed through problem reduction by imposing constraints on the size of each subset to a fixed value of $k$. In this context, for the full set $U$ with $n$ elements, each subset $S_j \subset U$ is defined such that $\left| S_j \right | = k$. Consequently, by uniformly, independently, and with replacement selecting $m$ subsets from the possible $C_n^k$ subsets, a random model akin to $F(n, m, k)$ can be formulated. Moreover, even in situations where constructing or analyzing the random model proves challenging, experimental approaches can be employed to obtain feasible approximations.
	
	\section{Conclusion}\label{sec_conclu} 
	
	This paper presents an efficient method for parameter setting of QAOA. The efficiency of the method is rooted in two primary aspects. Firstly, a preprocessing step, guided by the statistical properties of the specific problem's random model, is employed for the normalization of the Hamiltonian. Building on this, a robust initialization is obtained without incurring any optimization cost or pre-computation. Secondly, leveraging the analysis of the continuity of optimal adiabatic passage, the original parameter space $(\gamma, \beta)$ is transformed into $(\theta, \tau)$ to enhance the smoothness between parameters in adjacent layer of QAOA. Subsequently, an adiabatic-passage-based parameter setting method grounded in the continuity between optimal parameters is proposed, potentially exhibiting logarithmic optimization cost concerning $p$. Specifically, this method is applied to 3-SAT with an analysis of the random $k$-SAT model. The performance on random instances of 3-SAT aligns with the analysis, showcasing a superior advantage in parameter adjustment for QAOA.
	
	
	Opportunities for methodological enhancement exist, such as exploring alternative parameterizations for optimal adiabatic passage and applying other methods to utilize the continuity. 
	Furthermore, additional avenues for research beyond the scope of this method are identified as follows. Firstly, a more comprehensive exploration of the complexity of random 3-SAT can be undertaken, building upon the analyses and methodologies presented in this work. Secondly, applications can be extended to problems beyond the realm of combinatorial optimization.
	
	\section*{Acknowledgements}
	We acknowledge support by Innovation Program for Quantum Science and Technology (Grant Nos. 2021ZD0302900) and Natural Science Foundation of Jiangsu Province, China (Grant No.BK20220804).
	
	\bibliographystyle{quantum}
	
\end{document}